\documentclass[aps,letterpaper,nofootinbib,preprint,showpacs,preprintnumbers,amsmath,amssymb]{revtex4}%
\usepackage{latexsym}%
\usepackage[hypertex]{hyperref}%

\def\tR{{\widetilde R}}

\newcommand{\beq}{\begin{equation}}
\newcommand{\beqn}{\begin{equation}\nonumber}
\newcommand{\eeq}{\end{equation}}
\newcommand{\bea}{\begin{eqnarray}}
\newcommand{\bean}{\begin{eqnarray}\nonumber}
\newcommand{\eea}{\end{eqnarray}}

\begin{document}

\begin{center}
{\bf {\Large A Rotating, Inhomogeneous Dust Interior
\vskip 3mm
for the BTZ Black Hole}}

\bigskip
\bigskip

{{Cenalo Vaz\footnote{email: Cenalo.Vaz@UC.Edu} and
K. R. Koehler\footnote{email: Kenneth.Koehler@UC.Edu}}}

\medskip

{\it RWC and Department of Physics,\\ 
University of Cincinnati}\\
{\it Cincinnati, Ohio 45221-0011, USA}

\end{center}
\bigskip
\bigskip
\medskip

\centerline{ABSTRACT}
\bigskip
\medskip\noindent
We present exact solutions describing rotating, inhomogeneous dust with generic initial data
in 2+1 dimensional AdS spacetime and show how they are smoothly matched to the 
Banados-Teitelboim-Zanelli (BTZ) solution in the exterior. The metrics, which are the 
rotational analogues of the 2+1 dimensional LeMaitre-Tolman-Bondi (LTB) family, are described by 
their angular momentum and one additional constant which, together with the angular momentum, 
determines the energy density of the dust cloud. The weak energy condition gives a constraint 
on the angular momentum profile inside the cloud. Solutions can be stationary or time 
dependent, but only the time dependent solutions can be matched consistently to a BTZ 
exterior. No singularity is formed in either the stationary or the time dependent cases.
\medskip

%%%%%%%%%%%%%%%%%%%%%%%%%%%%%%%%%%%%%%%%%%%%%%%%%%%%%%%%%
\bigskip

\section{Introduction\label{intro}}

Exact solutions in general relativity are valuable tools to explore the range of possible 
behaviors allowed by Einstein's field equations. They are also useful as comparisons for 
numerical or approximate analytical solutions. Among these, solutions describing the gravitational 
collapse of matter are of considerable interest.

During the collapse of a neutral body, one expects the spacetime outside it to relax into the 
spacetime of a Schwarzschild or, more generally, a Kerr black hole, while the matter itself 
undergoes continual collapse until a singularity, either covered or locally naked, of spacetime 
forms. However, the matter may also avoid the singular region entirely either halting in a 
self-sustaining object, dissipating or, when rotation is present, even passing through a white 
hole into another universe. To examine the range of possibilities exact solutions describing black 
hole interiors with various forms of (preferably rotating) matter are essential.

Spherical, non-rotating gravitational collapse has been studied with various forms of matter 
and the indication  is that all the outcomes listed above may in principle be realized for 
regular initial data \cite{colSUM}. Critical behavior in the parameters, $p$, of the initial data have 
also been discovered \cite{christodoulou,choptuik93,evans93,gundlach97,choptuik00}: above some 
critical value of the parameter a singularity is formed, below it the matter eventually dissipates 
and at it the solutions are oscillatory \cite{wang01}. Of special interest is the fact that the 
critical values appear to be independent of spherical symmetry or the type of matter considered.

Compared to the spherical case, solutions with rotating matter smoothly matched to an exterior 
stationary vacuum are still poorly understood, although for some restricted classes of perfect 
fluids stationary solutions can be obtained \cite{stephani88,garfinkle97} by using a generalization 
of the solution generation procedure of \cite{ehlers59,harrison68,geroch71}.  The problem with obtaining 
general solutions is essentially in the complexity of the field equations.

A simpler setting for studying possible effects of rotation is the stationary solution of 
Banados, Teitelboim and Zannelli (BTZ) in 2+1 dimensions. The BTZ solution \cite{bhtz93,btz94}
is given in stationary coordinates as
\beq
ds^2 = - f(r) dt^2 + \frac{dr^2}f - r^2 (d\phi - \frac{J^2}{r^2}dt)^2,
\eeq
where
\beq
f(r)  = \Lambda r^2 - M + \frac{J^2}{r^2}
\eeq
and $M$ and $2J$ are conserved charges associated with the Killing vectors $\partial_t$ and 
$\partial_\phi$ They are interpreted as the mass and angular momentum of the black hole 
respectively (we take $8\pi G=1$). The solution can be obtained by non-standard identifications 
in AdS space \cite{bhtz93,vw94}, and there is no curvature singularity. There are, however, 
two horizons at which the Lapse vanishes,
\beq
r_\pm^2 = \frac{M}{2\Lambda}\left(1\pm \sqrt{1-\frac{4\Lambda J^2}{M^2}}\right),
\label{introbtz}
\eeq
an outer one at $r_+$ and an inner (Cauchy) horizon at $r_-$. Closed time-like 
curves exist wihin the Cauchy horizon, but there is evidence \cite{steif94} that the 
Cauchy horizon is unstable by mass inflation. It is of interest to know the fate of 
matter in the interior of the BTZ black hole and in particular if these features survive for 
reasonable initial data. For example the static BTZ black hole admits just one horizon and 
no curvature singularity, but when a collapsing and non-rotating dust interior is matched 
to the BTZ exterior a singularity does form at the center \cite{mannross93,gutti05}. 
Circularly symmetric null fluid collapse was examined by Husain \cite{husain95}, who found a variety of 
static limits, including the BTZ black hole but also other ``hairy'' black holes, depending 
on the equation of state. Self-similar perfect fluids were analyzed by Hirschmann et. al.
\cite{hirschmann04} with interesting results showing both naked and covered singularities. Chan 
et. al \cite{chanmann06,chanmann04} found a rotating interior containing null dust and were able 
to show explicitly that mass inflation \cite{poisisrael89} causes the inner (Cauchy) horizon to 
become unstable.  In this paper, we obtain exact, time dependent interiors representing rotating, 
time-like dust. Our solutions can be thought of as the rotational analogues of the 
LeMaitre-Tolman-Bondi [LTB] models in 2+1 dimensions.

Another, somewhat different motivation for exact solutions describing gravitational 
collapse stems from the fact that quantum gravitational effects in the vicinity of gravitational 
singularities are of fundamental interest. In a couple of recent publications \cite{vaz07,vaz08}, 
we developed a canonical, midisuperspace quantization of 2+1 dimensional collapse without rotation 
with interesting results. Not only was it possible to obtain Hawking radiation, along with 
grey-body factors that result from relaxing the near horizon approximation, but canonical 
quantization also provided a new and transparent interpretation of the entropy of the BTZ black 
hole: the black hole is viewed 
as a single shell formed by the collapse of many dust shells, each of which occupies one of the 
energy levels available to the single final-state shell. The energy spectrum of the single shell 
in the final state as obtained from canonical quantization coincides with the spectrum obtained in 
\cite{str98}, where it was proposed that 
because the asymptotic symmetry group of 2+1 dimensional gravity with a negative cosmological 
constant is generated by two copies of the Virasoro algebra \cite{brhe86}, its degrees of 
freedom should be described by two Conformal Field Theories (CFT's) at infinity. Explictly counting
the microstates of the black hole in the canonical theory yields an Area Law which, when compared 
to the Bekenstein-Hawking entropy, connects the boundary term at the center with the 
central charge of the AdS/CFT approach. This connection is potentially far reaching, but it 
was established only in the absence of rotation. In order to fully compare the results of the 
AdS/CFT approach and the midisuperspace quantization program it is essential to quantize the system 
with rotation. To carry this out, exact interior solutions are required.

The plan of this paper is as follows. In Section II we obtain the equations of motion for rotating 
dust collapse in 2+1 dimensions in the comoving frame. Einstein's equations determine both the 
velocity and the spatial gradient of 
the radius of collapsing shells, while also providing an integrability condition 
that relates the energy function and the angular momentum. We solve this integrability condition 
and examine the restrictions imposed by the weak energy condition on the angular momentum profile. 
In section 
III we solve the equations of motion for the dust shells. There are (critical) stationary solutions 
and time-dependent solutions. The time-dependent solutions are oscillatory and, in both cases, the 
evolution does {\it not} lead to the formation of a singularity. In Section IV, we address the matching 
conditions. Because the interior solutions are obtained in a comoving (hence corotating) frame, it 
is not possible to directly match them to the BTZ solution as given in \eqref{introbtz}, which 
is given in stationary coordinates. Therefore, we first obtain the transformations that take the 
stationary BTZ system to a comoving system. The transformations are analogous to those obtained for 
non-rotating black holes in \cite{ll}. Our derivation is based on an exact solution of the 
geodesic equations, given in Appendix A. This allows us to directly compare the interior metric 
representing the dust filled spacetime with the exterior vacuum (BTZ) metric. We show that the 
stationary interiors cannot be matched to the BTZ vacuum unless the ADM mass is negative. They 
may, however, serve as 2+1 dimensional cosmologies. We conclude in Section V with a few comments 
concerning the formation of trapped surfaces. Throughout, we follow the conventions of Weinberg 
\cite{weinberg}.

\section{Rotating Dust Ball in 2+1 AdS Gravity}

We consider a general axially symmetric metric in 2+1 dimensions of the form 
\beq
ds^2 = e^{2A}(dt-C d\varphi)^2 - e^{2B} dr^2 - (R^2+C^2 e^{2A})d\varphi^2,
\label{ansatz}
\eeq
where $A$, $B$, $C$ and $R$ are regarded as functions of $(t,r)$. The function 
$R(t,r)$ represents the curvature radius of the cloud. In the presence of a negative 
cosmological constant, Einstein's equations are of the form
\beq
G_{\mu\nu} + \Lambda g_{\mu\nu}=-8\pi G_3 T_{\mu\nu},
\label{einstein}
\eeq
where $G_3$ is the 2+1 dimensional Newton constant, $G_{\mu\nu}$ is the Einstein tensor, 
$-\Lambda$ ($\Lambda >0$) is the cosmological constant and $T_{\mu\nu}$ is the stress energy 
tensor (in what follows we set $8\pi G_3=1$). We take the metric in \eqref{ansatz} to be 
sourced by dust, so
\beq
T^{\mu\nu} = \varepsilon(t,r) U^\mu U^\nu,
\eeq
where $\varepsilon(t,r)$ represents the energy density of the collapsing dust cloud. 

\subsection{Comoving Frame}

In comoving coordinates, the only non-vanishing component of $T^{\mu\nu}$ is $T^{00}=
\varepsilon(t,r)$. The spatial components of the conservation equations then imply that
\bea
e^{2(A-B)}A'=0\cr\cr
C \dot A + \dot C = 0,
\label{bianchi1}
\eea
where dots refer to a derivatives with respect to $t$ and primes to derivatives with respect 
to $r$. The first of \eqref{bianchi1} requires that $A=A(t)$ is independent of $r$ 
and the second may be solved in terms of one integration function 
of $r$,
\beq
C(t,r) = K(r) e^{-A(t)}.
\eeq
It is possible to gauge fix so that $A=0$ making the metric in \eqref{ansatz} 
of the form
\beq
ds^2 = dt^2 - 2K(r) dt d\varphi - e^{2B(t,r)} dr^2 - R^2(t,r) d\varphi^2.
\label{comoving1}
\eeq
The metric describes rotating dust with angular velocity
\beq
\Omega(t,r) = -\frac{K}{R^2}
\eeq
and two non-vanishing components, $\omega_{r\varphi}=-\omega_{\varphi r}=-K'$, of the vorticity.
We assume that both $K$ and $K'$ are non-vanishing throughout the cloud except perhaps at the 
boundary. Uniform rotation occurs when $K'=0$; it is straightforward to show that the only solution 
of time-like dust with uniform rotation is the vacuum (BTZ) solution (this appears not to be the 
case for null dust \cite{chanmann06}). The time component of the conservation equations then reduces to
\beq
\partial_t \left(\varepsilon e^{B}\sqrt{K^2+R^2}\right) = 0.
\eeq
Its solution determines the energy density of the dust,
\beq
\varepsilon(t,r) = \frac{F'(r) e^{-B}}{\sqrt{K^2+R^2}},
\label{energydensity}
\eeq
in terms of the metric functions and one unknown integration function, $F(r)$, whose gradient is 
required to be positive by the weak energy condition. 

\subsection{Field Equations}

With \eqref{comoving1} we find that
\beq
G_{tr}+\frac{G_{r\varphi}}K = 0\Rightarrow (K^2+R^2)\dot B - R \dot R =0,
\eeq
which equation is solved by
\beq
e^{2B} = \frac{R^2+K^2}{W^2},
\eeq
where $W=W(r)$ is a function only of $r$, which we take to be greater than zero.\footnote{
One could work with a dimensionless label coordinate, 
\beqn
\rho = \int \frac{dr}W
\eeq
but we have chosen to keep $r$ and hence $W(r)$ arbitrary.}

The angular and radial components of Einstein's equations now give an expression for the 
acceleration, 
\beq
G^{\varphi\varphi}=\frac{G^{rr}}{W^2} = 4\Lambda (K^4+R^4)-W^2K'^2+4R^3 \ddot R
+4K^2 (2\Lambda R^2 +\dot R^2 + R\ddot R) = 0,
\eeq
which may be integrated and its solution given in terms of one integration function $E=E(r)$ 
as
\beq
{\dot R}^2 = -\Lambda R^2+(E-2\Lambda K^2)-\frac{4\Lambda K^4+K'^2W^2-4K^2 E}{4R^2}.
\label{Rdot}
\eeq
Inserting this solution into $G^{t\varphi}=0$ simplifies the equation to give a condition 
on the spatial gradient, $R'(t,r)$,
\beq
G^{t\varphi}=0\Rightarrow R'=R\left(-\frac{EK}{W^2K'}+\frac{W'}{2W}+\frac{K''}{2K'}\right)
+\frac 1R\left(-\frac{EK^3}{W^2K'}-\frac 12 KK'+\frac{K^2W'}{2W}+\frac{K^2K''}{2K'}\right).
\label{Rprime}
\eeq
As both $\dot R$ and $R'$ are determined by the field equations, there must be one integrability 
condition that enforces consistency. This is indeed the case and it is provided by the $(t,r)$ 
component of the field equations which gives a Riccati equation for $E$,
\beq
G^{tr}=0\Rightarrow 2E^2K+W^2K'(E-\Lambda K^2)'-EW(WK')'=0.
\label{gtr}
\eeq
(This condition can also be obtained by equating appropriate derivatives of \eqref{Rdot} and 
\eqref{Rprime}.) It can be viewed as prescribing $E$ in terms of the metric functions $W$ and $K$ up to 
a constant of integration. When $K'\neq 0$, the equation can be solved by quadratures 
since one of its solutions is $E=\sqrt{\Lambda}WK'$. The general solution turns out to be
\beq
E=\sqrt{\Lambda} WK'\coth\left(2\sqrt{\Lambda} \int \frac KW~ dr + H_0\right),
\label{Esol}
\eeq 
where $H_0$ is an arbitrary constant. So the particular solution we started with is the 
$H_0\rightarrow \infty$ limit of the general solution. The limit $H_0\rightarrow -\infty$ 
gives another particular solution, $E=-\sqrt{\Lambda} WK'$. 

The time-time component of the equations determines the energy density of the dust ball. Directly 
comparing $G^{tt}=-\varepsilon(t,r)$ with \eqref{energydensity} gives
\beq
F'(r) = -\frac{2E^2K^2}{W^3 K'^2}+\frac{2E +2\Lambda K^2}W - \left[\frac{(WK')'}{2K'}\right]',
\label{Fprime}
\eeq
which the weak energy condition requires to be positive. First viewing $F'$ as a function of $E$ we 
conclude that $E$ must lie between the roots of the equation $F'=0$, both of which must be real. 
Then taking into account \eqref{Esol}, we find that this amounts to a fairly complicated constraint 
on the angular momentum profile $K$ and the value of $H_0$. Equation \eqref{Fprime} can be formally 
integrated using the integrability condition in \eqref{gtr} and we obtain $F(r)$,
\beq
F(r)=\frac{E K}{WK'} - \frac{(WK')'}{2K'} + \int^r \frac EW~ dr,
\label{massfn}
\eeq
so long as $K'\neq 0$. We will show in section IV that consistent matching requires $K'$ 
to vanish at the boundary of the matter with the BTZ vacuum. As $K'$ approaches zero, 
\eqref{gtr} gives  
\beq
E=\frac{W^2 K''}{2K},
\label{kp0}
\eeq 
and only the last term in \eqref{massfn} survives in the limit. One can now explicitly 
verify that all the independent field equations are satisfied. 

In summary, the solutions are completely determined by the angular momentum and one 
arbitrary constant. The weak energy condition provides a complicated relationship 
between the angular momentum, its derivatives and the constant $H_0$, constraining the shape 
of the angular momentum profile. In the following section we integrate the remaining equation 
\eqref{Rdot} for $R(t,r)$.

\section{Exact Solutions}

The Ricci scalar depends linearly on the energy density, so a curvature singularity must
form when $R^2+K^2=0$. The Kretschmann scalar also diverges only in this limit, therefore there
is no good reason to terminate the spacetime at $R=0$ and one should be able to continue the 
solution into the region $-K^2<R^2<0$. This suggests that $R^2=0$ is just a coordinate singularity 
and that $R^2+K^2=0$ should be interpreted as a ring singularity. However, the azimuthal Killing 
vector $\xi=\partial_\varphi$ becomes timelike when $R^2<0$. Since its orbits must be periodically 
identified, closed time-like curves will exist in a neighborhood of the ring singularity.

We choose the negative square root in \eqref{Rdot} (to describe collapse) and integrate the 
equation, thereby getting 
\beq
\frac 12\int \frac{dx}{\sqrt{-\Lambda x^2 + A x + \frac B4}} = - t + \frac{Q(r)}{2\sqrt{\Lambda}},
\eeq
where we have set $x=R^2$, 
\beq
A=E-2\Lambda K^2,~~ B=4\Lambda K^4 +(WK')^2-4EK^2,
\eeq
and $Q(r)$ is an arbitrary function or $r$, which should be compatible with \eqref{Rprime}. 
By transforming to $y=x+K^2$, it becomes clear that real solutions exist only so long as 
$E\geq \sqrt{\Lambda} |WK'|$, and we find\footnote{This is also indicated by the solution
\eqref{Esol} of the integrability condition \eqref{gtr}.} 
\beq
R^2+K^2 = \frac{E}{2\Lambda} - \frac{\sqrt{E^2-\Lambda W^2 K'^2}}{2\Lambda}
\sin(2\sqrt{\Lambda} t - Q).
\label{Rsol}
\eeq
Thus $R^2+K^2$ is strictly positive at all times and no strong curvature singularity will form.  
The solution $E=\sqrt{\Lambda} |WK'|$ of \eqref{gtr}, obtained in the $H_0\rightarrow \infty$ limit,
determines a stationary (time-independent) solution. In this case, the dust is in a self-sustaining 
distribution with vanishing radial velocity, but for the weak energy condition to hold it is 
necessary for 
\beq
K'> \frac 1{2\sqrt{\Lambda}}\left[\frac{(WK')'}{2K'}\right]'
\eeq
to be verified throughout the dust cloud. 

For all finite values of $H_0$ the solutions are oscillatory. Equation \eqref{Rprime} may be satisfied 
only if $Q(r)$ is an arbitrary constant, $Q_0$. This constant then determines the physical radius of 
any shell at the initial time, say $t=0$
\beq
R^2(0,r) = \frac{E-2\Lambda K^2}{2\Lambda} + \frac{\sqrt{E^2-\Lambda W^2 K'^2}}{2\Lambda}
\sin(Q_0),
\label{Rinitial}
\eeq
and reflects a limited freedom in our choice of initial scaling, {\it i.e.,} the shell 
radius $R(t,r)$ at the initial time. For instance, if we take $Q_0=0$ and use 
\eqref{Rinitial} to reexpress the function $F(r)$ in terms of the initial energy density 
profile, we find
\beq
F(r) = \int^r \varepsilon(0,r') \frac{K^2+R^2}{W}\vert_{t=0,r'}dr'
= \frac 1{2\Lambda} \int^r \varepsilon(0,r') \frac{E(r')}{W(r')}dr'.
\eeq
Furthermore, the initial radial velocity profile is obtained from \eqref{Rdot},
\beq
v_0^2(r)={\dot R}^2(0,r)= \frac{E^2-\Lambda W^2K'^2}{2(E-2\Lambda K^2)}.
\eeq
Thus, if the dust cloud begins with zero initial radial velocity then $E=\sqrt{\Lambda}
WK'$ and the dust continues in stationary flow. For a general $E$ (compatible with 
the positive energy condition),
\beq
R^2(t,r) = \frac{E-2\Lambda K^2}{2\Lambda}\left[1-\frac{\sqrt{2}v_0}{\sqrt{E-2
\Lambda K^2}}\sin(2\sqrt{\Lambda}t)\right]
\label{Rsol2}
\eeq
reexpresses the solution \eqref{Rsol} in terms of the initial velocity profile, $v_0(r)$.

\section{Matching to the BTZ exterior}
 
In the previous section we obtained collapsing dust solutions with rotation in 
the comoving system. For the solutions to represent the interior of a black hole, 
they should be shown to go over smoothly to a stationary exterior vacuum spacetime. 
In 2+1 dimensions with a negative cosmological constant, this is given by the BTZ 
solution in \eqref{introbtz}, which we write here as
\beq
ds^2 = f(\tR) dT^2-\frac{d\tR^2}{f(\tR)} - \tR^2 \left(d\phi-\frac{J}{\tR^2} dT
\right)^2,
\eeq
where
\beq
f(\tR)=\Lambda \tR^2 -  M + \frac{J^2}{\tR^2}.
\eeq
$T$, $\tR$ and $\phi$ are stationary coordinates, and it is not possible to 
directly compare the angular parts of the comoving dust solutions 
as is usually done in the absence of rotation. Instead we first use the solutions of the 
geodesic equations to set up a comoving coordinate system for the BTZ metric and then 
compare the interior and exterior in the usual way. 

\subsection{Comoving Coordinates for the BTZ vacuum}

The equations for time-like geodesics of the BTZ space-time can be integrated exactly 
in terms of two constants of the motion, $P$ and $L$, representing the energy and angular 
momentum of the orbits respectively \cite{cruz94}. For our purposes we need only the 
velocities (see Appendix A). With proper time as affine parameter, they may be expressed as
\bea
&&U^T=\frac 1{\sqrt{f}}\cosh\eta,\cr\cr
&&U^R=\sqrt{f}\sinh\eta\cos\alpha,\cr\cr
&&U^\phi=\frac 1R\sinh\eta\sin\alpha + \frac J{\tR^2\sqrt{f}}\cosh\eta,
\label{geodesics}
\eea
where
\bea
&&\cosh\eta=\sqrt{\frac Pf} - \frac{JL}{\tR^2\sqrt{f}},\cr\cr
&&\sin\alpha = \frac L{\tR\sinh\eta}.
\label{etalpha}
\eea
Consider a system of coordinates $(t,r,\varphi)$ that are comoving with respect to these 
orbits; in this system, $U^t=1$ and $U^\varphi=U^r=0$. Assuming that the transformation 
between the canonical BTZ coordinates and this comoving system depends only on 
$(t,r)$, we write 
\beq
\left(\begin{matrix}
dT\cr\cr
d\tR\cr\cr
d\phi\end{matrix}\right) = \left(\begin{matrix}
\frac{\cosh\eta}{\sqrt{f}} & T' & 0\cr\cr
\sqrt{f}\sinh\eta\cos\alpha & \tR' & 0\cr\cr
\frac 1R\sinh\eta\sin\alpha+\frac{J}{\tR^2\sqrt{f}}\cosh\eta & \phi' & 1
\end{matrix}\right)\left(\begin{matrix}
dt\cr\cr
dr\cr\cr
d\varphi\end{matrix}\right).
\label{trans1}
\eeq
Because $P$ and $L$ are constant along geodesics, they may be viewed as functions 
of $r$. There is therefore some freedom in our choice of the comoving coordinates. 
We may fix this freedom by imposing coordinate conditions in such a way as to obtain 
the simplest form for the BTZ metric in the comoving system. For example,
requiring $g_{tr}=0$ we find 
\beq
T' = \frac{\tR'}f\tanh\eta\cos\alpha,
\label{condition1}
\eeq
but we must verify that the condition is compatible with the integrability
of the function $T(t,r)$. From \eqref{condition1},
\beq
T(t,r) = h(t) + \int \frac{d\tR}f\tanh\eta\cos\alpha
\eeq
and therefore
\beq
\dot T = \dot h + \frac{\dot \tR}f\tanh\eta\cos\alpha \Rightarrow 
\frac{\cosh\eta}{\sqrt{f}} = \dot h+\frac{\sinh^2\eta\cos^2\alpha}{\sqrt{f}\cosh\eta},
\eeq
where we have used \eqref{trans1}. Inserting $\eta$ and $\alpha$ from the solutions 
in \eqref{etalpha} we find that $\dot h(t)$ is independent of $r$ if and only if 
$P(r)=1$ and $L(r)=-J$. This gives $\dot h = 1$ and therefore
\beq
t = T -\int \frac{d\tR}f\tanh\eta\cos\alpha.
\eeq
This choice of $P$ and $L$ also ensures that 
\beq
\phi' = \frac{J\tR'}{\tR^2 f}\tanh\eta\cos\alpha
\label{condition2}
\eeq
is compatible with the integrability of $\phi$, for we have
\beq
\phi = \varphi + g(t)+\int \frac{Jd\tR}{\tR^2 f}\tanh\eta\cos\alpha \Rightarrow 
\dot\phi = \dot g + \frac{J\dot \tR}{\tR^2 f}\tanh\eta\cos\alpha,
\eeq
and using \eqref{trans1} together with \eqref{etalpha} we easily determine 
$\dot g\equiv 0$. Thus
\beq
\varphi=\phi -  \int \frac{Jd\tR}{\tR^2 f}\tanh\eta\cos\alpha.
\eeq
With \eqref{condition1} and \eqref{condition2}, the transformation from stationary to 
comoving coordinates becomes
\beq
\left(\begin{matrix}
dT\cr\cr
d\tR\cr\cr
d\phi\end{matrix}\right) = \left(\begin{matrix}
\frac{\cosh\eta}{\sqrt{f}} & \frac{\tR'}f\tanh\eta\cos\alpha & 0\cr\cr
\sqrt{f}\sinh\eta\cos\alpha & \tR' & 0\cr\cr
\frac 1R\sinh\eta\sin\alpha+\frac{J}{\tR^2\sqrt{f}}\cosh\eta & \frac{J\tR'}
{\tR^2 f}\tanh\eta\cos\alpha & 1
\end{matrix}\right)\left(\begin{matrix}
dt\cr\cr
dr\cr\cr
d\varphi\end{matrix}\right)
\label{trans2}
\eeq
and the BTZ metric can be expressed in comoving coordinates with the line element 
\beq
ds^2 = dt^2 + 2Jdtd\varphi-\frac{R'^2}{1+J^2/\tR^2}dr^2 - 
\tR^2d\varphi^2.
\label{comovingBTZ}
\eeq
Note that
\beq
{\dot R}^2 = f\sinh^2 \eta \cos^2\alpha = -\Lambda \tR^2 + (1+M-\Lambda J^2)
+ \frac{J^2(1+M)}{\tR^2}
\label{BTZeqmot}
\eeq
has the same structure as \eqref{Rdot}, with constant coefficients determined by $M$ 
and $J$. Its solution can be given as
\beq
R^2(t,r)=\frac{1+M-\Lambda J^2}{2\Lambda}-\frac{1+M+\Lambda J^2}{2\Lambda}\sin[2\sqrt{\Lambda}
t-Q(r)],
\eeq
where $Q(r)$ remains arbitrary. It will be seen that $R^2(t,r)$ is bounded from above 
but not from below. Thus the comoving coordinates cover only a portion of the spacetime.
This happens because there exists a radial upper bound for the geodesics of massive particles
\cite{cruz94}. Nevertheless, the upper bound for $R(t,r)$  is larger than 
the radius of the outer horizon and \eqref{comovingBTZ} is sufficient for verifying 
the matching conditions between the interior and the exterior spacetimes.

\subsection{Matching Conditions}

Direct comparison between the first fundamental forms describing the interior in 
\eqref{ansatz} and the exterior in \eqref{comovingBTZ} on the boundary $r=r_b$,  
shows that one may identify the coordinates $t$ and $\varphi$ along with $K(r_b)=
K_b=-J$ and the physical radius $R(t,r_b)=\tR(t,r_b)\stackrel{\text{def}}{=} R_b(t)$. 
Comparing the expressions for the radius function, we find that $K_b'=K'(r_b)=0$,
\beq
1+M +\Lambda J^2 = E_b
\label{Eb}
\eeq
and $Q(r_b)=Q_0=0$. Because the vorticity at the boundary is vanishing, there is 
an infinite discontinuity in the density $F'(r)$ there, but $F$ itself
remains continuous across the boundary.

We must also ensure that the second fundamental forms match. The second fundamental 
form of the boundary surface in the interior has three non-vanishing components, {\it viz.,}
\beq
K^\text{in}_{t\varphi}=K^\text{in}_{\varphi t} = \left.\frac{WK'}{2\sqrt{K^2+R^2}}\right|_{r_b},~~ 
K^\text{in}_{\varphi \varphi} = \left.\frac{WRR'}{\sqrt{K^2+R^2}}\right|_{r_b},
\eeq
and in the exterior there is only one non-vanishing component, 
\beq
K^\text{out}_{\varphi\varphi} = \sqrt{J^2+R_b^2}.
\eeq
Clearly, $K_b'=0$ as before and 
\beq
R_b' = \frac{J^2}{W_b R_b}+\frac{R_b}{W_b}.
\eeq
This is guaranteed in the limit as $r\rightarrow r_b$ if $E_b =  W_b^2 K_b''/2K_b$ which,
as shown in \eqref{kp0} solves the integrability condition \eqref{gtr} with $K_b'=0$. We 
conclude  that while the vorticity must vanish, $K''(r)$ cannot vanish at the boundary and 
moreover that $K''_b/K_b>0$ when the BTZ mass is positive. 

On the other hand we had obtained stationary interiors when $E$ takes the critical 
value $E=\sqrt{\Lambda} W K'$. Evidently, then $E_b=\sqrt{\Lambda} W_b K_b'=0$ for the stationary 
interiors and, by \eqref{Eb}, these metrics cannot be consistently matched to the BTZ vaccuum 
except when the ADM mass is negative. They may, however, serve as inhomogeneous, rotating 2+1 
dimensional cosmologies. 

\section{Concluding Remarks}

Much can be said about the nature of the sigularity formed in gravitational collapse by 
examining the expansion, $\Theta$, of a congruence of twist free null geodesics \cite{tpsingh97}. 
However, if one is interested only in the trapped surfaces ($\Theta=0$) there is a simpler 
but equivalent approach which we now briefly illustrate for the vacuum given in the time dependent 
form of \eqref{comovingBTZ}. Consider null geodesics in this spacetime. The existence of an 
azimuthal Killing vector field, $\xi=\partial_\varphi$, implies a locally conserved quantity
\beq
\xi_\mu U^\mu = g_{\varphi t} U^t + g_{\varphi\varphi} U^\varphi = L,
\label{conservedL}
\eeq
where $U^\mu = dx^\mu/d\lambda$ is tangent to a geodesic and $L$ may be interpreted as the angular 
momentum carried by it. Therefore, according to \eqref{conservedL}, a congruence of null geodesics 
carrying zero angular momentum (such a congruence is twist free and hypersurface forming) satisfies
\beq
U^\varphi=\frac{J}{R^2} U^t,
\eeq
so the null conditon,
\beq
\left(1+\frac{J^2}{R^2}\right){U^t}^2 -\frac{R'^2}{1+J^2/R^2} {U^r}^2  - R^2 
\left(U^\varphi-\frac{J}{R^2}U^t\right)^2 = 0,
\eeq
requires that along these geodesics
\beq
\frac{dt_n}{dr} = \pm \frac{R'}{1+J^2/R^2},
\label{nulltreqn}
\eeq
where the positive sign refers to outgoing geodesics and the negative sign to infalling ones. 
The radius along an outgoing null ray is then given by $R_n=R(t_n(r),r)$, where $t_n(r)$ is 
the integral curve of \eqref{nulltreqn}, and the horizons are determined as the extrema of $R_n$,
\beq
\frac{dR_n}{dr}=0={\dot R}_n \frac{dt_n}{dr} + R_n'.
\eeq
Substituting \eqref{nulltreqn} for outgoing radial geodesics and using \eqref{BTZeqmot}, we 
find the condition
\beq
R_n'\left(\frac{{\dot R}_n}{1+J^2/R_n^2}+1\right) = R_n'\left(-\frac{\sqrt{-\Lambda R_n^2 + 
1+M-\Lambda J^2+\frac{J^2(1+M)}{R_n^2}}}{1+J^2/R_n^2}+1\right) = 0.
\eeq
Assuming that $R_n'\neq 0$, the condition becomes
\beq
-\Lambda R_n^2 + 1+M-\Lambda J^2 +\frac{J^2(1+M)}{R_n^2} = \left(1+\frac{J^2}{R_n^2}\right)^2,
\eeq
which admits three real solutions (provided that $M>4\Lambda J^2$), two of which are positive,
\beq
{R_n^\pm}^2 = \frac{M}{2\Lambda}\left[1\pm \sqrt{1-\frac{4\Lambda J^2}{M}}\right].
\eeq
These will be recognized as the inner and outer horizons of the BTZ black hole, as determined 
in the introduction from the vanishing of the Lapse function in the stationary frame. The third
root, $R_n^2=-J^2$, may be ignored as there is no reason to extend the vacuum solution to negative 
values of $R^2$ \cite{bhtz93}.

We may apply the same reasoning to determine the formation of an apparent horizon in the dust 
filled region. With $U^\varphi=-K/R^2$, the null condition requires that along these twist free 
null geodesics
\beq
\frac{dt_n}{dr} = \pm \frac{\sqrt{R_n^2+K^2}}W.
\label{nulltreqn2}
\eeq
The turning points of an outgoing ray are then given by
\beq
0 = \frac{dR_n}{dr} = \left(\frac{\dot R_n\sqrt{R_n^2+K^2}}{W}+R_n'\right),
\eeq
and thus we find that the condition for the formation of an apparent horizon is
\beq
-{\dot R}_n \sqrt{R_n^2+K^2}=WR_n',
\eeq
where $\dot R_n$ is given in \eqref{Rdot} and $R_n'$ in \eqref{Rprime}. 

For the stationary solutions, this is just the condition $R'=0$, {\it i.e.,} at shell crossings. 
For the time dependent solutions, the condition once again yields a cubic equation for $R_n^2$ 
and therefore admits at least one real root. Unfortunately, while the equation itself is 
straightforward to solve, the solutions are difficult to analyze without making further assumptions 
on the shape of the angular momentum profile in the dust ball. This goes beyond the scope of the 
present work.

In this paper we have examined time-like dust collapse with rotation in 2+1 dimensions. 
A simplification of Einstein's equations was achieved by going to the comoving coordinate 
system. We found that all solutions are determined by one function ($K$, the angular momentum) and 
one arbitrary constant ($H_0$), and we examined the limits on $K$ and $H_0$ imposed by the weak 
energy condition. We determined stationary as well as oscillatory, time dependent solutions, but 
no singularity formation. To match the solutions in the comoving system to the BTZ black hole,
we found a comoving system for the black hole, explicity giving the transformations from the 
stationary to the comoving frames. We demonstrated that the time dependent solutions can be matched 
to a BTZ exterior with positive ADM mass provided that $K_b''/K_b>0$ at the boundary, but the 
stationary solutions cannot be so matched. Thus the stationary spacetimes may be thought of as 2+1 
dimensional, inhomogeneous cosmologies, but not as black hole interiors. We have also presented a 
simplified approach to addressing the issue of trapped surfaces, although the conditions imposed by 
Einstein's equations are not as transparent as they are in the case of circular collapse and we could
recover little information from the solutions without specializing to particular cases. However, 
the solutions do not at first glance seem to make a good toy model for analytically examining 
critical behavior because there are not varied outcomes and so it is not clear how useful such 
specializations might prove.

The cases of zero and positive cosmological constant (dS) may also be addressed from our 
expressions, in the first case by taking the limit as $\Lambda\rightarrow 0$ and in the second case 
by analytic continuation.
\bigskip\bigskip

\noindent{\bf Acknowledgments}
\medskip

\noindent We are grateful to Louis Witten and T.P. Singh for stimulating discussions and 
valuable comments.
\bigskip\bigskip

\centerline{\bf Appendix A}

In this appendix we obtain the geodesics given in \eqref{geodesics} and \eqref{etalpha} for 
the BTZ spacetime.  The condition for being time-like,
\beq
f{U^T}^2 - f^{-1}{U^\tR}^2 - \tR^2 \left({U^\phi}-\frac{J}{\tR^2}{U^T}\right)^2 = 1,
\eeq
allows us to reexpress the four velocities in terms of a ``boost'' and a 
rotation according to,
\bea
U^T &=& \frac 1{\sqrt{f}} \cosh\eta\cr\cr
U^\tR &=& \sqrt{f} \sinh\eta \cos\alpha\cr\cr
{U^\phi}-\frac{J}{2\tR^2}{U^T} &=& \frac 1\tR \sinh\eta \sin\alpha
\label{parameterization}
\eea
The geodesic equations can be written as
\bea
&&\frac{dU^T}{ds} +\frac{d\ln f}{ds} U^T - \frac{2J}{f\tR}\left({U^\phi}-\frac{J}{\tR^2}{U^T}\right)
U^\tR =0\cr\cr
&&\frac{dU^\tR}{ds} +\frac{f'}{2}\left(f{U^T}^2-\frac{{U^\tR}^2}f\right)-\tR f\left({U^\phi}^2
-\frac{J^2{U^T}^2}{\tR^4}\right)=0\cr\cr
&&\frac{dU^\phi}{ds} - \frac{2J^2}{f\tR^3}\left(U^\phi-\frac{J}{\tR^2}U^T\right)U^\tR + 
\frac{2U^\tR}{\tR} \left(U^\phi + \frac{J(\ln f)'}{\tR} U^T\right)=0
\eea
Subtracting the first from the last, we arrive after some manipulation at the equation
\beq
\frac d{ds}\left(U^\phi-\frac{J}{\tR^2}U^T\right)+\frac 2\tR\left(U^\phi-\frac{J}
{\tR^2}U^T\right)U^\tR=0,
\eeq
which is solved by
\beq
U^\phi-\frac{J}{\tR^2}U^T = \frac{L}{\tR^2},
\label{LL}
\eeq
where $L$ is constant on geodesics. Inserting this into the equation for $U^T$, we find 
that it can be integrated to yield
\beq
U^T = \frac{\sqrt{P}}f - \frac{JL}{f\tR^2}
\eeq
where $P$ is also constant on geodesics. Combining this with \eqref{parameterization}
\beq
\cosh \eta = \sqrt{\frac Pf} - \frac{JL}{\tR^2\sqrt{f}}
\eeq
and, using \eqref{LL} in \eqref{parameterization}, 
\beq
\sin\alpha = \frac L{\tR\sinh\eta}
\eeq

\end{document}